\documentclass[journal=jacsat,manuscript=article]{achemso}
\usepackage{caption}
\usepackage{graphicx}
\usepackage{wrapfig}
\usepackage[font=small,skip=2pt]{caption}
\usepackage{xcolor}
\usepackage{array}
\usepackage{varioref}
\usepackage{hyperref}
\usepackage{subfig}
\usepackage{cleveref}
\usepackage{lineno,hyperref}
\usepackage{csquotes}
\usepackage{diagbox}
\usepackage{makecell}
\usepackage{soul}
\usepackage{setspace}
\usepackage{mwe}
\usepackage{adjustbox}
\usepackage{float}
\usepackage{multirow}

\author{Sharath Kumar C}
\affiliation[1]{School of Physics, IISER Thiruvananthapuram, Vithura, Thiruvananthapuram-695551, India}
\author{Amal Sebastian}
\affiliation[1]{School of Physics, IISER Thiruvananthapuram, Vithura, Thiruvananthapuram-695551, India}
\author{Athira S}
\affiliation[1]{School of Physics, IISER Thiruvananthapuram, Vithura, Thiruvananthapuram-695551, India}
\author{Ranjit Singh}
\affiliation[1]{School of Physics, IISER Thiruvananthapuram, Vithura, Thiruvananthapuram-695551, India}
\author{Akhil Chakravarthy Kakarlamudi}
\affiliation[2]{School of Chemistry, IISER Thiruvananthapuram, Vithura, Thiruvananthapuram-695551, India}
\author{Andrews P. Alex}
\affiliation[1]{School of Physics, IISER Thiruvananthapuram, Vithura, Thiruvananthapuram-695551, India}
\author{Sivaranjana Reddy Vennapusa}
\affiliation[2]{School of Chemistry, IISER Thiruvananthapuram, Vithura, Thiruvananthapuram-695551, India}
\author{D. Jaiswal-Nagar}
\affiliation[1]{School of Physics, IISER Thiruvananthapuram, Vithura, Thiruvananthapuram-695551, India}
\email{deepshikha@iisertvm.ac.in}
\title[An \textsf{achemso}]
{\textbf{One-dimensional magnetism in a facile spin 1$/$2 Heisenberg antiferromagnet with a low saturation field}}

\date{}
\begin{document}
\maketitle

\begin{abstract}
This work reports the synthesis, structure and magnetic properties of the single crystal of a facile spin 1$/2$ one dimensional Heisenberg antiferromagnet bis(4-aminopyridinium) bis(oxalato)cuprate(II) dihydrate, (C$_5$H$_7$N$_2$)$_2$[Cu(C$_2$O$_4$)$_2$].2H$_2$O. Single crystals of large sizes of the title compound were obtained using the technique of liquid-liquid diffusion or layer diffusion with 100 $\%$ yield. Single crystal X-ray diffraction measurements revealed a very good quality of the grown single crystals with a small value of goodness of fit R obtained at 1.058. The structure comprises corner sharing CuO$_6$ octahedra resulting in Cu-Cu chains in the a-direction that are very well isolated in the $b$ and $c$ directions. Density functional theory (DFT) with three different basis sets (B3LYP/6-311++G(d,p); B3LYP/LanL2DZ and B3LYP/6-311++G(d,p), B3LYP/LanL2DZ) generated the optimized geometry of a monomeric unit as well as its vibrational spectra. The vibrational frequency corresponding to the CuO$_6$ octahedron was found in the experimentally obtained IR spectrum that matched very well with the theoretically obtained IR spectra incorporating the mixed basis. Temperature dependent dc magnetic susceptibility revealed a low temperature peak suggesting the presence of low dimensional magnetism in the system. Bonner-Fisher fit confirmed the one dimensional nature of the magnetic interaction with an exchange coupling constant of 1.23 K. Magnetisation measurements along with Quantum Monte Carlo simulations confirm this metal-organic crystal to be a very good spin-1/2 Heisenberg antiferromagnet with a low saturation field H$_s$ of 1.75 T.

\end{abstract}
\section{Introduction}
Metal ions linked by coordination ligands form a metal organic coordination polymer (MOCP), wherein, a large array of polymers is made by smaller monomeric units comprising the metal ions linked with coordination ligands \cite{battern,morsali,blake}. Such MOCP's can vary from simple one-dimensional chains with small ligands to large mesoporous frameworks like metal–organic framework \cite{james,yaghi}. These materials find very diverse potential applications including magnetism \cite{carlin,kahn,meundaeng,xu} , optics \cite{wang,fleisch,blanco}, sensors \cite{lee,lu}, medical diagnostics \cite{huh,gao}, energy storage and batteries \cite{sakunth}, data storage \cite{sun,schmid} etc. Of the many applications mentioned above, MOCP's find a very important application as molecular magnets in the area of low dimensional magnetism both from a fundamental as well as an applied point of view where they are now being searched as potential candidates for quantum computers \cite{mathewprr,sahling}.\\
In recent times, there has been a tremendous impetus on research in low dimensional magnetism since fluctuations from equilibrium properties tend to be highly enhanced in lower spatial dimensions resulting in exotic phenomenon like quantum phase transitions \cite{vojta,haldane,affleck}. Lower the spatial dimension, higher is the effect of quantum-mechanical fluctuations. So, one dimensional (1-D) magnets in which the spatial dimension of consideration is one, is a very important research area in low dimensional magnets. 1-D molecular magnets comprise long molecular chains whose central magnetic atom interact readily with each other along the chain via the bridging ligands, but are insulated from each other by side groups or co-groups \cite{kahn,birgeneau}. The strength of MOCP's lies in the ease with which the bridging ligands and the side ligands can be tuned \cite{birgeneau,battern}, thus, tuning the strength and sign of the magnetic exchange interaction between the magnetic metal centres.\\ 
A spin 1$/$2 antiferromagnetic Heisenberg chain \cite{lieb,affleck} (AfHc) is the seminal model of quantum many-body systems whose ground state maps onto Tomonaga-Luttinger liquid comprising entangled spins \cite{lieb,affleck,haldane,bogoliubov,eggert}. A spin 1$/$2 AfHc is also quantum critical with respect to an applied magnetic field where it has a line of critical points till the saturation field H$_s$, above which it transforms to a fully polarised state \cite{wolf,mathewprr,sahling,affleck}. The inorganic systems Sr$_2$CuO$_3$ and SrCuO$_2$ which are good representations of spin 1$/$2 AfHc have very high exchange coupling constant, J$/$k$_B$ of the order of 2000 K resulting in saturation fields of the order of 2000 T, limiting the field dependent measurements that can be done in laboratory conditions. So, in order to be able to study the details of the fascinating properties exhibited by a spin 1$/$2 AfHc, it is desirable to have systems with moderate values of exchange coupling constant in the range of few Kelvins so that the corresponding saturation feild is of the order of few Teslas.\\
In this work, we report the synthesis of single crystals of a facile spin 1$/$2 AfHc system, (C$_5$H$_7$N$_2$)$_2$[Cu(C$_2$O$_4$)$_2$].2H$_2$O (abbreviated as CuD henceforth), with a moderate value of exchange coupling constant J$/$k$_B$ = 1.23 K and saturation field H$_s$ = 1.75 T. The structure has been reported previously in \cite{pan} but no other details of the system was given. Our single crystals were grown using the technique of liquid-liquid diffusion that resulted in the growth of large sized single crystals with 100 $\%$ yield. The facile system is a MOCP in which the layers of [Cu(C$_2$O$_4$)$_2$]$^{-2}$ ions connected by long axial Cu–O bonds result in Cu-Cu chain in the crystallographic a-direction formed by corner sharing oxygen atoms of the CuO$_6$ octahedra. The vibration frequency corresponding to the CuO$_6$ octahedra was found in the experimental FTIR spectrum. Density functional theory incorporating B3LYP hybrid functional and three different basis sets, namely, (i) 6-311G+(d,p), (ii) LanL2DZ and (iii) 6-311++G(d,p), LanL2DZ (ONIOM model) were used to theoretically obtain the FTIR spectra. It was found that the ONIOM model containing the metal centre as well as the organic ligands matched the experimentally obtained FTIR spectrum the best, with the identification of the frequency corresponding to the CuO$_6$ octahedron vibration. dc magnetic susceptibility as well as magnetisation measurements reveal that the system is an excellent realization of a spin-1$/$2 antiferromagnetic Heisenberg chain confirmed by Bonner-Fisher fits as well as Quantum Monte-Carlo simulations that fit the experimentally obtained curves very well.

\section{Experimental Section}
\subsection{Materials and General Methods}
All the reagents required for growing the single crystals of CuD were purchased from commercial sources and used as received; 4-aminopyridine (Merck-Aldrich, 99\%), Potassium bis(oxalato) cuprate(II)dihydrate (Merck-Aldrich, 99.999\%), Ethyl acetate(Merck-Aldrich, 99.8\%). The crystals were grown in Memmert's incubator (Model No. IPP30 PLUS 32L). Single-crystal X-ray diffraction (SCXRD) measurements were done on a Bruker APEX II CCD diffractometer using graphite-monochromatized Mo-K$_\alpha$ radiation ($\lambda =0.71073 \AA$) at room temperature 296(2) K. IR spectra (3983–400 $cm^{-1})$ were recorded using Shimadzu’s spectrometer (Model No. IRPrestige-21) using the KBr pellet technique. Magnetization measurements in the temperature range of 1.9 K to 300 K were performed on a single crystal of mass 14 mg on Quantum Design’s SQUID (Superconducting QUantum Interference Device) magnetometer (Model MPMS3). Lower temperature measurements in the range of 0.39 K to 2 K were done on a Helium 3 insert attached to MPMS3 (Model iHelium3). In all the experiments, the crystals were cooled in the zero
field (ZFC cooling) mode to the lowest possible temperature after which a field of 100 Oe was applied and the data collected in the warming up mode.

\subsection{Synthesis}
\begin{figure}[h]
	\begin{center}
		\includegraphics[width=1\textwidth]{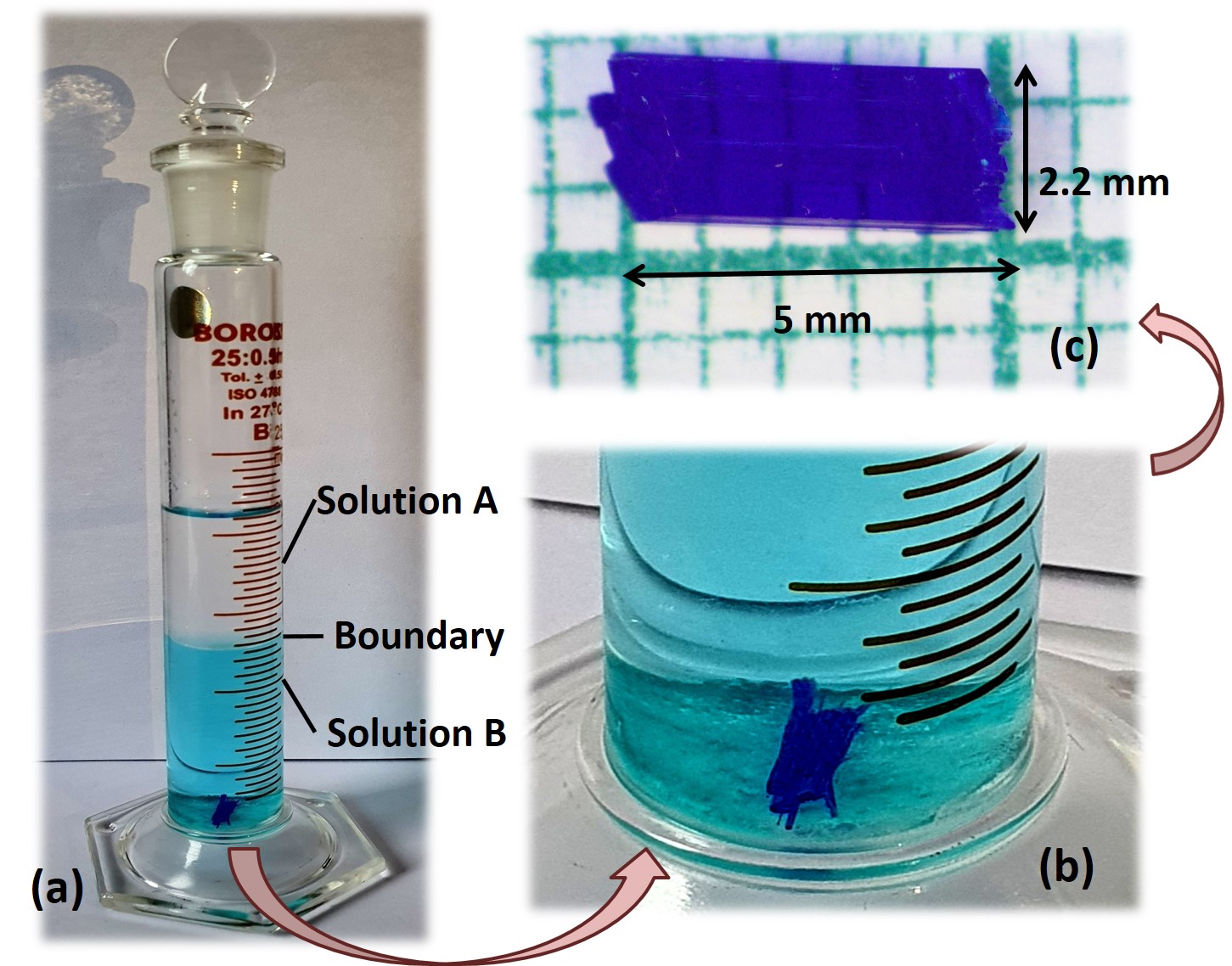}
		\caption{(colour online) (a) Solution A forming the bottom layer and Solution B forming the top layer of the liquid-liquid diffusion technique. Crystal growth happens at the boundary of the two layers. (b) Large sized CuD crystal at the bottom of the cylinder. (c) Extracted single crystal on a graph paper for size comparison.}
		\label{fig:synthesis}
	\end{center}
\end{figure}

The present study involves the growth of single crystals of CuD using the technique of liquid/liquid diffusion which involves the slow diffusion of one solution into another \cite{adam,kitayama}. In this technique, two solutions are added in a container such that they form distinct layers. This is possible when the two solutions have different densities. A solution in which the target compound is soluble is taken as the solution A and forms the bottom layer. A lesser dense solution in which the target compound is insoluble is taken as solution B and forms the top layer. Solution B, then, diffuses into the solution A very slowly creating an oversaturated solvent mixture at the boundary that has a lower solubility for the target compound than the pristine solution A. With the passage of time, as more and more solution B diffuses into solution A, the solubility of the mixture at the interface decreases more and more for the target compound until the oversaturated solute is forced to precipitate at the liquid-liquid boundary where the crystals form \cite{adam,kitayama}. The type of solvents used for making the two solutions determine the structure and morphology of the obtained crystals.\\
For initiating the crystal growth using the liquid-liquid diffusion technique, 0.1 g (0.3 mmol) of potassium bis(oxalato) cuprate(II) dihydrate was dissolved in 12 ml of distilled water to make the solution A in a beaker. Similarly, the lesser dense solution B (which doesn't dissolve in solution A) was made by dissolving 0.059 g (0.6 mmol) of 4-Aminopyridine in 12 ml of ethyl acetate in another beaker. To make the two layers for the liquid-liquid diffusion technique, solution A was added to the bottom of a cylindrical glass vessel of about 2 cm diameter and 25 ml capacity as shown in Fig. \ref{fig:synthesis} (a) forming the bottom layer. Solution B from the beaker was slowly dribbled through the wall of the inclined vessel to ensure that no drops fell directly into the solution A to disturb the liquid/liquid boundary. Due to the difference in the densities of solution A and solution B, solution A remains as the top layer (see Fig. \ref{fig:synthesis} (a)). The cylindrical vessel with the two layers was kept inside an incubator at 25$^\circ$C for crystal growth. After 45 days, a needle was formed at the liquid-liquid boundary around which tiny crystallites formed. With the passage of time, the tiny crystallites diffused towards the needle increasing its size. After another 15 days, the large single crystal that formed into the boundary fell to the bottom of the cylinder as shown in Fig. \ref{fig:synthesis} (a) and Fig. \ref{fig:synthesis} (b) on an expanded scale. The formed crystal was filtered out and rinsed with a minimal amount of cold distilled water. The remaining excess solvent was removed by blotting the crystal on a piece of clean tissue paper. The large single crystal was cut to an approximate rectangular shape for other measurements. The liquid-liquid diffusion technique gave us 100 \% yield with exceptionally large sized crystals that were at least 5 mm $\times$ 2 mm $\times$ 2 mm in size as shown in Fig. \ref{fig:synthesis} (c) and 35 mg weight.  \cite{assmus}.

\subsection{X-ray data collection and structure determination}
A tiny crystal with dimensions 0.250 x 0.120 x 0.050 mm$^3$ was mounted in random orientation on a glass fibre for SCXRD measurements. Intensity distribution indicated a monoclinic structure with the space group as P21$/$c. This choice was confirmed by a successful refinement. Unit cell parameters as well as orientation matrix were found by least-squares treatment of the setting angles of $1580$ reflections, all of which were independent, in the $2.6 \leq 2\theta \leq 25^{\circ}$ range. The data were reduced to structure factors in the usual manner, corrected for absorption, transformed and averaged in the required symmetry. Atomic positions were located by the direct methods using the programme SHELXT \cite{shelxt}. The structure was, then, refined on F$^2$ by full-matrix least-square techniques using SHELXL program package \cite{shelxl}. The refinement was carried out by using 1$/$$\sigma_2$ weights. Only reflections with F $> 2\sigma$ weights were included in the refinement. All nonhydrogen atoms were refined anisotropically. The positions of hydrogen atoms were added in an idealized geometrical positions. 10611 reflections were collected out of which out of which 1610 were independent. Table \ref{table:xrd} gives the crystallographic data and structural refinement parameters for CuD. Bond lengths and  bond angles are given in supplementary material. From the Table \ref{table:xrd}, it can be seen that the goodness of fit R has a low value of 1.058 indicating an extremely good quality of the grown crystal.

\begin{table}
\begin{center}
\begin{tabular}{|ccccc|}
  \hline 
 \bf {Emperical formula}&    \hspace{2cm} &\bf{C14 H18 Cu N4 O10} & &\\ 
 \hline
 Formula weight & \hspace{2cm} & 465.86 & &\\ 

 Crystal System & \hspace{2cm} & Monoclinic & &\\ 
 
 Space Group & \hspace{2cm} & P 21/c & &\\ 
  
Unit cell dimensions & \hspace{2cm} & a = 3.724(10) Å & \hspace{2cm} & $\alpha$ = 90$^\circ$\\ 
 
& \hspace{2cm} & b = 20.39(6) Å & \hspace{2cm} & $\beta$ = 90.51(3)$^\circ$ \\ 
 
& \hspace{2cm} & c = 11.97(3) Å & \hspace{2cm} & $\gamma$ = 90$^\circ$  \\  
  
Volume & \hspace{2cm} & 909(4)Å$^{3}$ & & \\ 
 
Z & \hspace{2cm} & 2 & & \\ 
  
Density (calculated) & \hspace{2cm} & 1.703 Mg/m$^3$ & & \\ 
 
Absorption coefficient & \hspace{2cm} & 1.267 mm$^{-1}$ & & \\
 
Index ranges & \hspace{2cm} & 1.703 Mg/m$^3$ & & \\
  
Goodness-of-fit on F$^{2}$ &  & 1.079 & & \\ 
  
Final R indices [I $>$ 2$\sigma$(I)]  & \hspace{2cm} & R1 = 0.0264, wR2 = 0.0630 & &\\ 
 
R indices (all data) & \hspace{2cm} & R1 = 0.0354, wR2 = 0.0673 && \\ 
 
Extinction coefficient & \hspace{2cm} & n/a & & \\
 
Largest diff. peak and hole & \hspace{2cm} & 0.222 and -0.292 e.$\AA ^{-3}$ & & \\
\hline
 \end{tabular}  
\caption{Crystallographic data and structure refinement of C$_5$H$_7$N$_2$)$_2$[Cu(C$_2$O$_4$)$_2$].2H$_2$O.}
\label{table:xrd}
\end{center} 

\end{table}

\subsection{Computational Details}
In order to create the optimised geometry of the monomeric form of CuD and estimate the various vibrational frequencies and bands, density functional theory (DFT) was employed using the Gaussian 16 program package without any constriant on the geometry \cite{gaussian16}. For these calculations, atomic co-ordinates obtained from the structure using SCXRD measurements were input to the Gaussian 16 package. DFT calculations were done at the level of "Becke, 3 parameter, Lee Yang Parr" (B3LYP) hybrid functionals that are very commonly used to approximate the exchange-correlation energy functional in DFT \cite{becke,lyp}. Since the crystal has a metal centre (Cu$^{2+}$) as well as connecting organic ligands (oxalate and amminopyridine groups), three different kinds of basis sets were used for the calculations: i) 6-311G+(d,p), ii) LanL2DZ, and iii) an ONIOM model with LanL2DZ for the Cu atom (metal center) and 6-311G+(d,p) for the rest of the atoms (organic part) \cite{abkari,fernando}. The charge and spin multiplicity in all quantum chemical computations are taken as 0 (neutral) and 2 (doublet), respectively. The computed vibrational modes were analyzed using Gaussain16 program package \cite{frisch}, which is used for molecular structural and vibrational analysis. The calculations converged to an optimized geometry since there were only real harmonic vibrational wavenumbers, revealing the localization of energy minima. Harmonic infrared vibrational wavenumbers were calculated for the fully optimized molecular geometry.\\
To calculate the field dependence of magnetisation of a spin 1$/2$ AfHc-a quantum many body system, Quantum Monte Carlo (QMC) computations were performed. The implementation of QMC was done using the directed loop stochastic series expansion (SSE) application written in Python code using Algorithms and Libraries for Physics Simulations (ALPS) package \cite{alps}. The calculations were done on a spin 1$/$2 antiferromagnetic Heisenberg chain containing 100 particles with maximum 50000 sweeps and 5000 thermalizations.

\section{Results and discussion}
\subsection{Crystal structure details}
\begin{figure}[h]
	\begin{center}
		\includegraphics[width=1\textwidth]{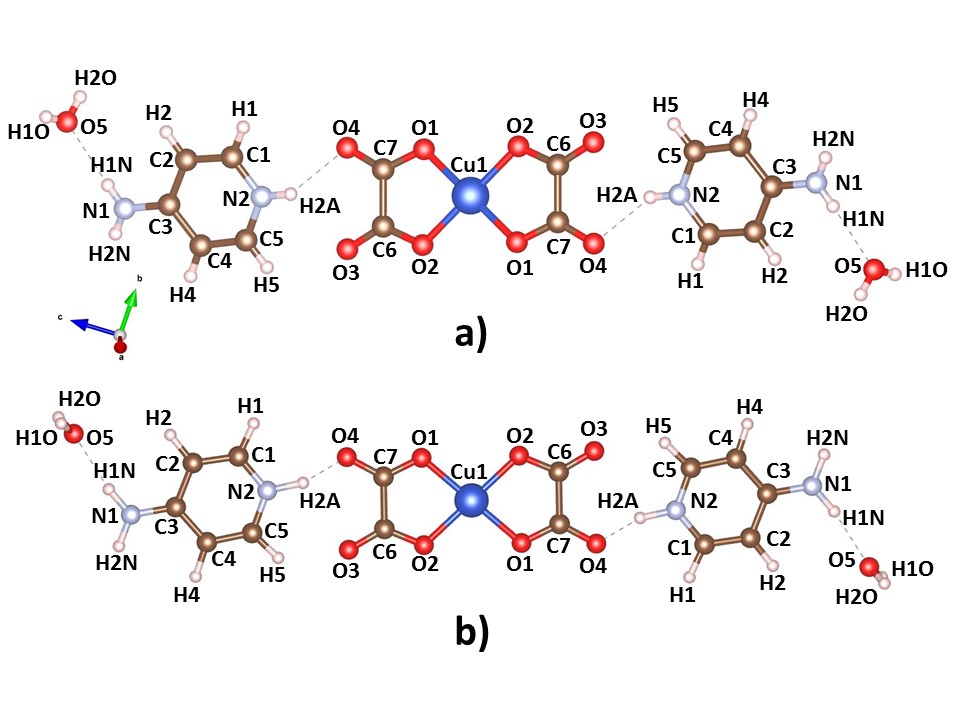}
		\caption{(colour online) (a) Basic structural unit of C$_5$H$_7$N$_2$)$_2$[Cu(C$_2$O$_4$)$_2$].2H$_2$O obtained from single crystal x-ray diffraction measurements with atoms numbered. (b) Optimised geometry of C$_5$H$_7$N$_2$)$_2$[Cu(C$_2$O$_4$)$_2$].2H$_2$O obtained from DFT calculations using B3LYP/6-311++G(d,p), B3LYP/LanL2DZ basis sets.}
		\label{fig:single molecule}
	\end{center}
\end{figure}

As shown in Fig.\ref{fig:single molecule} (a), the basic structural unit of CuD consists of one bis(oxalato)cuprate(II) complex anion- [Cu(C$_2$O$_4$)$_2$]$^{-2}$, two 4-aminopyridinium cations- [C$_5$H$_7$N$_2$]$^+$, and two lattice water molecules. The bis(oxalato)cuprate(II) complex ion is located at the centre of inversion at the origin such that Copper(II) ion is coordinated by two chelating oxalate ions. Here, the copper(II) ion is bonded  to four oxygen atoms of the two oxalato(2-) ligands (average Cu–O bond length: 1.9285 \AA) in the equatorial plane, thus making a 5 membered ring of Cu1-O2-C6-C7-O1 on either sides of Copper(II) ion. The two equitorial oxygen atoms O1 and O2 surround the Cu1 atom in a bidentate and symmetrical mode. The 4-aminopyridinium rings on either side of bis(oxalato)cuprate(II) complex are weakly coupled to the complex via weak hydrogen bonds O4/O5..H2A. There are two more hydrogen bonds in the basic structural unit that connect the water molecules with the 4-aminopyridinium rings in the form of H1N..O5 hydrogen bonds. From the Fig. \ref{fig:single molecule} (a), it can also be seen that the water molecules are \textit{trans} to each other according to the centrosymmetry.

\begin{table}
	\caption{Experimental and calculated (B3LYP/LanL2DZ, B3LYP/6-311++ G(d,p),B3LYP/LanL2DZ+6-311++
		G(d,p)) geometrical parameters of (C$_5$H$_7$N$_2$)$_2$[Cu(C$_2$O$_4$)$_2$].2H$_2$O}
	\label{table:parameters}
	\begin{tabular}{|c|c|c|c|c|}
		\hline 
		\bf{Parameters} &\bf{Experimental} & \bf{B3LYP/} & \bf{B3LYP/} & \bf{B3LYP/}\\ 
		& & \bf{LanL2DZ}& \bf{6-311++ G(d,p)} & \bf{LanL2DZ+6-311++}\\
		\hline
		\bf{Bond length[{\AA}]} &  & & & \\ 
		
		Cu1-O1 & 1.9287(14) & 1.981 & 1.970 & 1.984\\ 
		
		Cu1-O2 & 1.9279(14) & 1.960 & 1.950 & 1.964\\ 
		
		\bf{Bond angle($^\circ$)} &  &  & &\\ 
		
		O1-Cu1-O2 & 85.33(6) & 83.59 & 84.08 & 83.53 \\ 
		
		Cu1-O1-C7 & 112.89(14) & 113.79  &  112.82 & 112.99\\ 
		
		Cu1-O2-C6 & 112.56(12) & 114.91 & 113.97& 112.99\\ 
		\bf{Torsion angle($^\circ$)} & & & &\\
		O4-C7-O1-Cu1 & -175.50(18) & 179.98 & 180.00 & 180.00\\
		C6-C7-O1-Cu1 & 6.2(2) & -0.01 & 0 & 0\\
		C7-C6-O2-Cu1 & -0.2(2) & 0 & 0 & 0 \\
		\hline 
	\end{tabular} 
\end{table}

Fig. \ref{fig:single molecule} (b) shows the optimised geometry of the CuD molecule obtained from the DFT calculations with B3LYP functional and the dual ONIOM basis. From the figure, it can be seen that the optimised geometry has an excellent match with the experimentally obtained structural unit, where the bond lengths as well as bond angles match very well with each other. The only difference between experimentally and computationally obtained structures is in the dihedral angle of water molecules that are at an angle of 17 $^{\circ}$C with respect to the 4-aminopyridinium ring. The difference of this dihedral angle arises since the dihedral angle between the planes in which 4-aminopyridinium cation and bis(oxalato)cuprate(II) complex lies is about 17$^{\circ}$.
  
\begin{figure}[h]
	\begin{center}
		\includegraphics[width=1\textwidth]{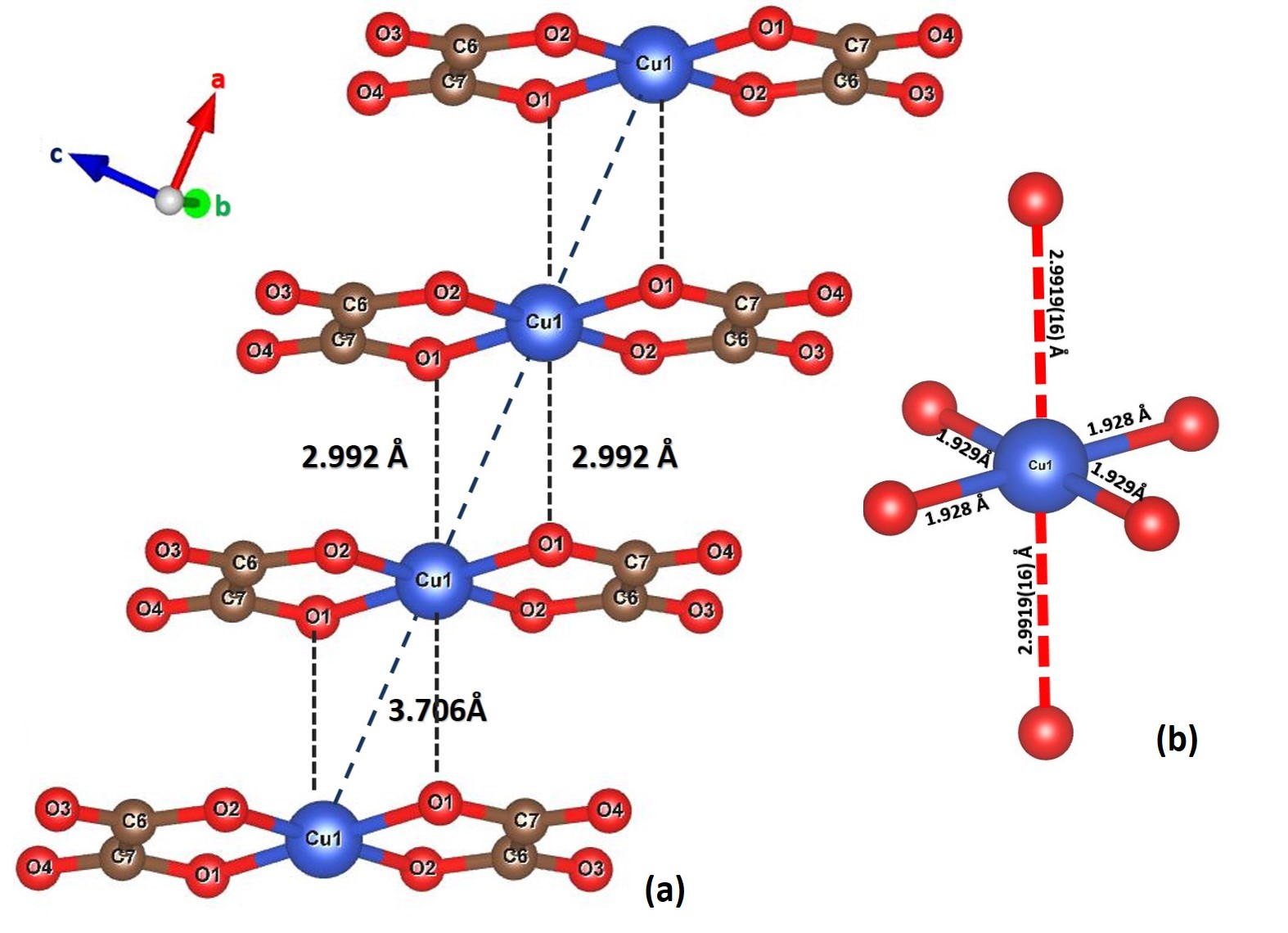}
		\caption{(colour online) (a) Chain structure of C$_5$H$_7$N$_2$)$_2$[Cu(C$_2$O$_4$)$_2$].2H$_2$O along the a-axis. (b) Oxygen environment of Cu$^{2+}$ ion in the crystal structure.}
		\label{fig:alonga}
	\end{center}
\end{figure}

Apart from the optimised geometry as shown in Fig. \ref{fig:single molecule} (b) using B3LYP/6-311++G(d,p), B3LYP/LanL2DZ basis sets, we also obtained optimised geometry of the structure generated by DFT using i) B3LYP/6-311++G(d,p) basis set and ii) B3LYP/LanL2DZ basis set 
as described in the computational details section. A comparison of few selected bond lengths, bond angles and dihedral angles are shown in Table \ref{table:parameters}. Complete information can be found in the supplementary material. From the Table \ref{table:parameters}, it can be observed that there is quite a good matching of the experimentally obtained parameters from SCXRD measurements and those obtained from DFT calculations employing different basis sets. Additionally, the parameters obtained from the dual basis B3LYP/6-311++G(d,p), B3LYP/LanL2DZ have the closest match with the experimentally obtained parameters. This is understandable since the dual basis has components from both the metal centre as well as the organic parts. It is noteworthy that the calculations done on a monomeric unit in the gas phase has such a good match with the experimentally obtained parameters which is a crystalline state of a many-body system. This suggests that the interactions between various monomeric units of the crystalline state is very weak.\\ 
Cu(II) coordinates with four O atoms in the basal plane to form a square planar geometry and the O2-Cu1-O1 bite angle is 85.33$^{\circ}$, a value which is similar to those found in previously reported metal complexes\cite{geiser,nenwa,keene,dazem}. The chelating oxygen atoms with the Copper atom are at an average distance of 1.92 \AA. The O1 atom within the plane deviates the most from a least square planar arrangement and is slightly displaced towards the next Copper ion of the chain to which it becomes semi-coordinated as shown in Fig. \ref{fig:alonga} (a). So, the first [Cu(C$_2$O$_4$)(4-apy)(H$_2$O)] unit is linked to the second [Cu(C$_2$O$_4$)(4-apy)(H$_2$O)] unit via an oxalate O1 atom which appears in the axial position of the second Cu1 atom where the Cu1-O1 bond length is 2.992 \AA. This results in a prolate distorted octahedral geometry of CuO$_6$ octahedra such that the coordination sphere of the copper ion is 4 + 2. The long Cu1-O1 bond lengths are indicative of weak interactions in the complex. It should be noted that the Cu1-Cu1 bond length is 3.706 \AA, a value similar to that obtained by Geiser et al. \cite{geiser}. So, the compound forms uniform stacks of [Cu(C$_2$O$_4$)$_2$]$^{-2}$ entities polymerised along the a-axis through symmetry-related O1 atoms, forming straight Cu(II) chain with regular spacing of Cu$\cdots$Cu = 3.706 \AA as shown in Fig. \ref{fig:alonga} (a).

\begin{figure}
	\begin{center}
		\includegraphics[width=0.9\textwidth]{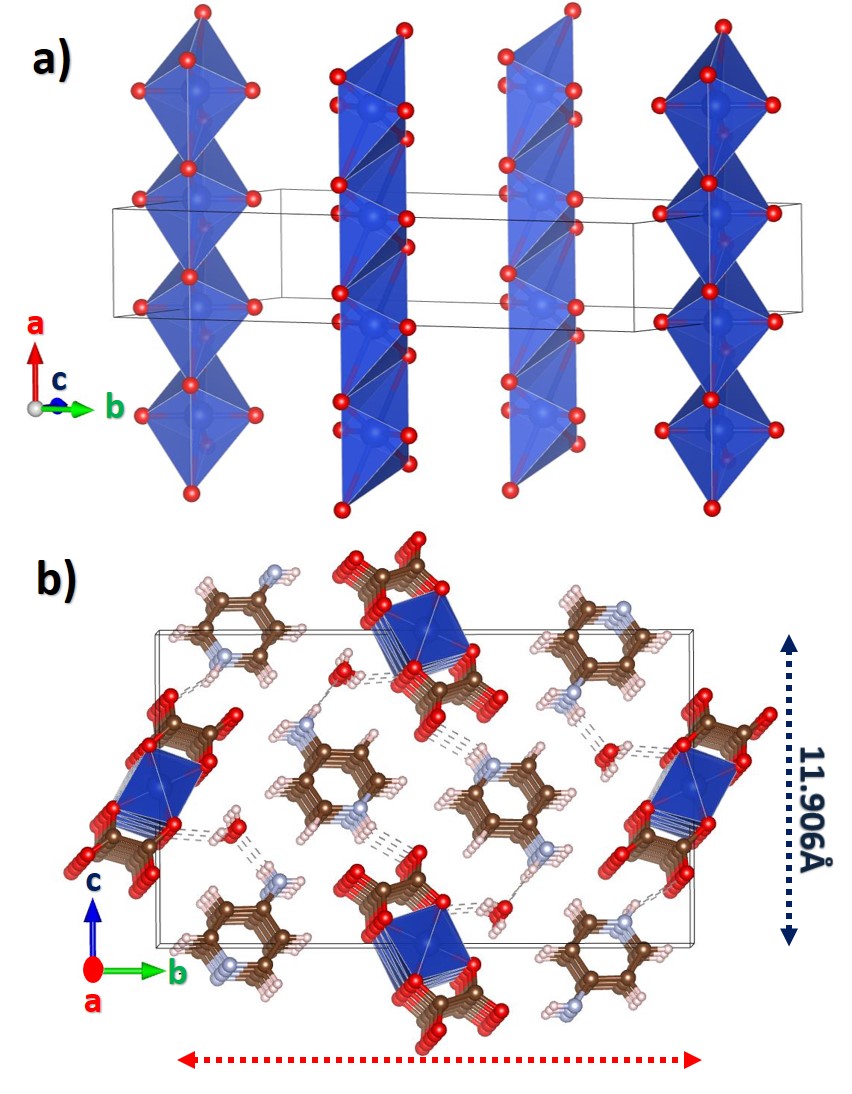}
		\caption{(colour online) Chain structure of C$_5$H$_7$N$_2$)$_2$[Cu(C$_2$O$_4$)$_2$].2H$_2$O along the bc-plane. Cu-Cu distance along b-axis is 20.297 \AA while that along the c-axis is 11.906 \AA.}
		\label{fig:alongbc}
	\end{center}
\end{figure}

The nearest oxygen environment of a copper atom in the crystal structure of CuD is shown in Fig. \ref{fig:alonga} (b). Each copper atom is surrounded by six nearest-neighbour oxygen atoms forming a distorted octahedron CuO$_6$. The apical Cu-O distance is the same for both the apical oxygen atoms at 2.992 \AA and is larger than the Cu-O distance within the plane (Cu-O1 = 1.9287 \AA and Cu-O2 = 1.9279 \AA).\\   
So, the crystal structure of CuD consists of infinite chains of Cu$^{2+}$O$_6$ octahedra sharing common apical corners as shown in Fig. \ref{fig:alongbc} (a), resulting in zig-zag chains along the a-direction. This corner sharing octahedron structure is similar to that obtained in vanadium phosphate structures like $\beta$-Na$_4$VO(PO$_4$)$_2$ \cite{panin}. The one dimensional chains containing the magnetic copper centres are well separated perpendicular to the a-axis by large distances. For instance in the $c$-direction, the chains are connected by  hydrogen bond linkages that form between hydrogen atoms of the pyridine ring and O atom of the water molecule of one [Cu(C$_2$O$_4$)$_2$(H$_2$O)] unit and O atom of the oxalate group of the other [Cu(C$_2$O$_4$)$_2$(H$_2$O)] unit as shown in Fig. \ref{fig:alongbc}. The Cu-Cu distance in the c-direction is 11.906 \AA. In the b-direction, the [Cu(C$_2$O$_4$)$_2$(H$_2$O)] units are decoupled from each other and lie at a large Cu-Cu separation of 20.297 \AA  from each other. From the above discussions, it is clear that CuD mimicks a quasi-one dimensional system of a chain running along the a-direction such that the interaction of the Cu(II) ions in the other two directions is negligible due to large interchain distances and screening by non
magnetic 4-aminopyradine in these directions.

\subsection{FTIR analysis}
\begin{figure}
	\begin{center}
		\includegraphics[width=0.9\textwidth]{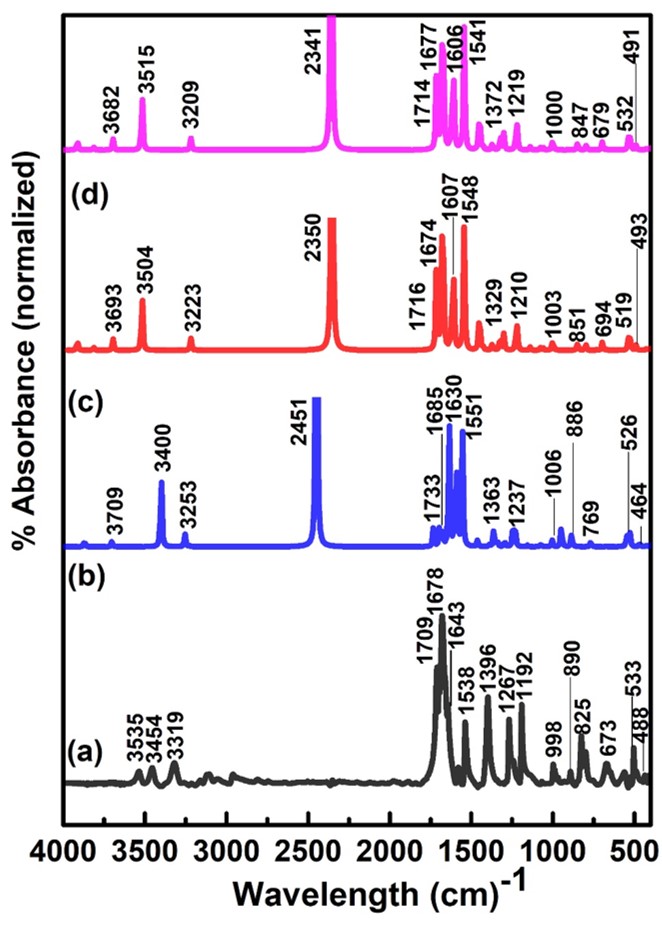}
		\caption{(colour online) (a) Experimental FTIR spectra of C$_5$H$_7$N$_2$)$_2$[Cu(C$_2$O$_4$)$_2$].2H$_2$O at room temperature. FTIR spectra of C$_5$H$_7$N$_2$)$_2$[Cu(C$_2$O$_4$)$_2$].2H$_2$O generated from DFT calculations using basis sets as (b) B3LYP/LanL2DZ, (c) B3LYP/6311G++(d,p), and (d) B3LYP/LanL2DZ, B3LYP/6-311++ G(d,p).}
		\label{fig:FTIR}
	\end{center}
\end{figure}

The above structure of corner sharing CuO$_6$ octahedron must have a signature in the IR spectra that must have a characteristic vibrational frequency corresponding to the CuO$_6$ octahedron. To confirm this, FTIR measurements were done on CuD at room temperature as shown in Fig.\ref{fig:FTIR}(a) (black solid curve). The IR spectrum exhibits three weak absorption bands at 3535 cm$^{-1}$, 3454 cm$^{-1}$ and 1643 cm$^{-1}$. By measuring the FTIR of 4-aminopyridine, Arnaudov et al. \cite{arnaudov} found IR peaks at 3507 cm$^{-1}$, 3413 cm$^{-1}$ and 1623 cm$^{-1}$ that were attributed to NH$_2$-stretching and scissoring- $\nu_{as}^f$(NH$_2$), $\nu_{s}^f$(NH$_2$) and $\delta^f$(NH$_2$) respectively. So, the first two IR bands are attributed to the stretching and scissoring vibrations of NH$_2$. The weak band at 3319 cm$^{-1}$ is attributable to $\nu_{(O-H)}$ of the water molecules. The IR spectrum also shows 3 strong absorption bands at 1709cm$^{-1}$, 1678cm$^{-1}$ and 1643cm$^{-1}$ followed by other lower intensity bands at lower frequencies. From the FTIR measurements on bis(oxalato)cuprate(II) dihydrate, two closely separated bands at 1640 cm$^{-1}$ and 1665cm$^{-1}$ were identified as arising due to $\nu_{asym(O-C=O)}$ vibrations of the molecule \cite{muraleedharan}. So, the vibrational frequencies at 1678cm$^{-1}$ and 1643cm$^{-1}$ indicate the deprotonation of the oxalate group \cite{nakamato,muraleedharan}. Similarly, the bands observed at 1267cm$^{-1}$ and 825 cm$^{-1}$ are attributed to $\nu_{sym(C-O)}$ + $\delta_{O-C=O}$ of the oxalate group while those at 998 cm$^{-1}$ are attributable to $\nu_{(C-C)}$ stretching vibrations \cite{nakamato,kripa,muraleedharan,edwards}. Finally, the absorption band at 503$^{-1}$ and 435 cm$^{-1}$ correspond to $\nu_{(Cu-O)}$ vibrations \cite{edwards}.\\ 
The experimental FTIR spectra of Fig. \ref{fig:FTIR} (a) confirms the vibrational frequencies of 4-aminopyridine, oxalate and water molecules present in the structure of CuD. The FTIR spectrum of Fig. \ref{fig:FTIR} (a) is also found to exhibit a frequency at 673 cm$^{-1}$ which is the characteristic frquency of CuO$_6$ octahedron \cite{su,raju}, thus, confirming the presence of CuO$_6$ octahedron in the crystal structure of CuD. FTIR spectra obtained by DFT analysis using 3 different basis sets (i) B3LYP/LanL2DZ, (ii) B3LYP/6-311++G(d,p), and (iii)
B3LYP/6-311++G(d,p), B3LYP/LanL2DZ (ONIOM)) are shown in Figs. \ref{fig:FTIR} (b), (c) and (d) respectively. From the graphs, it can be seen that the matching of the theoretically obtained spectra for describing the vibrational frequencies of 4-aminopyridine, oxalate and water molecules is better for the  B3LYP/6-311++G(d,p) basis set and the combination ONIOM basis when compared to the B3LYP/LanL2DZ basis. It is also to be noted that the characteristic vibration frequency of CuO$_6$ octahedron at $\sim$ 673 cm$^{-1}$ is absent from the FTIR spectra computed using the B3LYP/LanL2DZ basis since this basis doesn't incorporate the metal centre. However, between the B3LYP/6-311++G(d,p) basis and the mixed ONIOM basis, the matching of the ONIOM basis set which has a metal centre and an organic ligend, is better. It is also reasonable that a basis set that describes the organomettalic compound better would have a better matching with the experimentally obtained spectrum, thus demonstrating the usefulness of the ONIOM basis.  

\subsection{Magnetic Susceptibility and magnetisation}
\begin{figure}
	\begin{center}
		\includegraphics[width=1\textwidth]{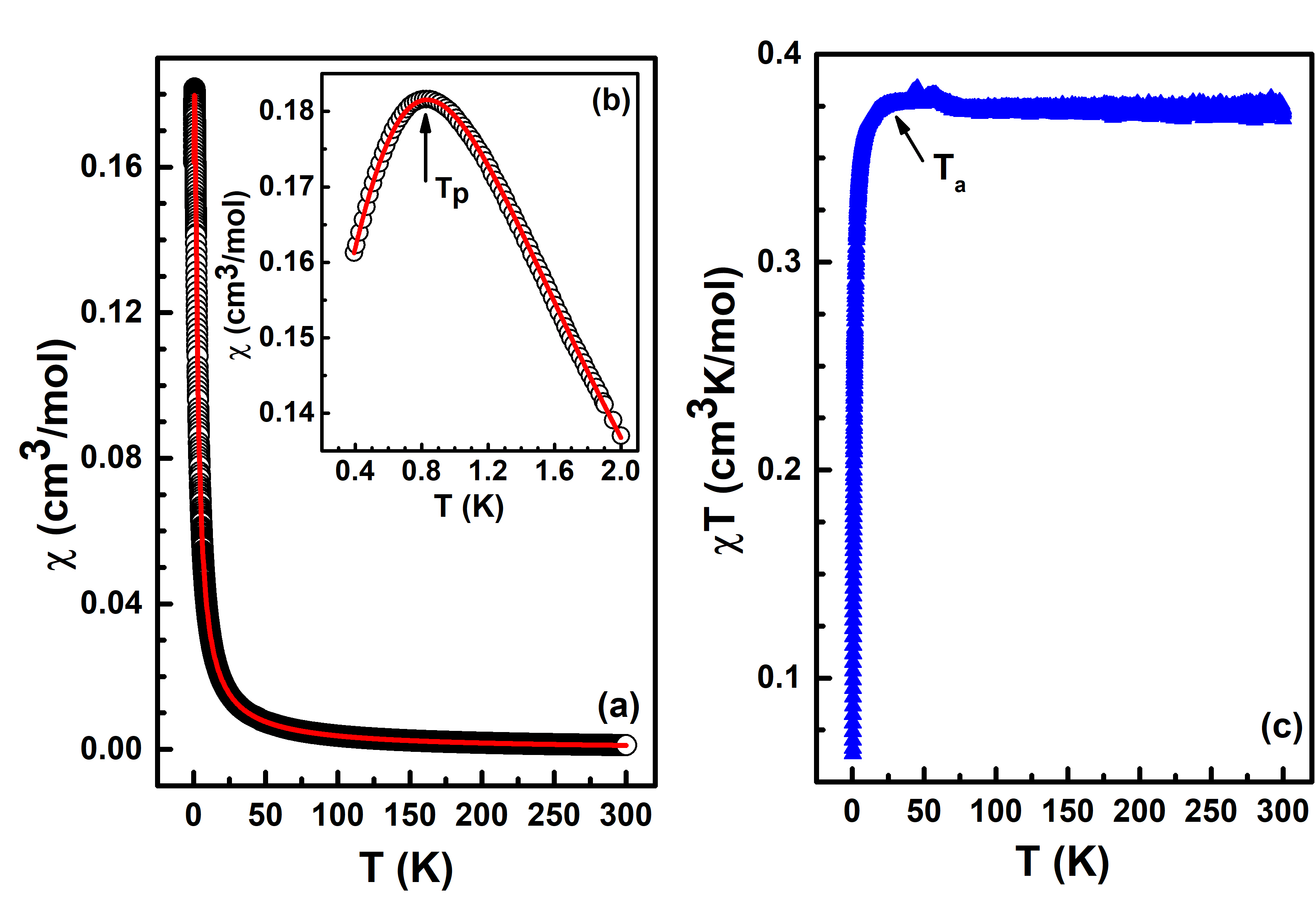}
		\caption{(colour online) (a) Temperature variation of dc magnetic susceptibility in C$_5$H$_7$N$_2$)$_2$[Cu(C$_2$O$_4$)$_2$].2H$_2$O at an applied field of 500 Oe. Open circles represent the data points while the red solid curve is a fit to the Bonner-Fisher model. See text for details. (b) $\chi$-T data on an expanded scale to show the low temperature peak clearly. (c) Temperature variation of $\chi$-T. }
		\label{fig:M-T}
	\end{center}
\end{figure}
To understand the magnetic ground state of CuD, a single crystal of CuD was subjected to magnetisation measurements. Open circles in Fig. \ref{fig:M-T} (a) represent the temperature variation of dc magnetic susceptibility $\chi$(T) = M(T)/H where M is the magnetisation and H is the applied field, measured at an applied field of 500 Oe where the direction of field application is parallel to the chain direction. The magnetic data was found to be reversible on heating and cooling with no indication of any hysteresis.\\ 
From Fig. \ref{fig:M-T} (a), it can be observed that with a decrease in temperature, $\chi$(T) increases smoothly till it reaches a maximum at $\sim$ 0.78 K $\pm$ 0.2 K, below which the dc magnetic susceptibility starts to decrease. The peak is clearly visible in Fig. \ref{fig:M-T} (b) where the peak temperature, T$_p$, is marked by an arrow as the temperature where d$\chi /$dT = 0. The low temperature peak is indicative of low dimensional magnetism \cite{eggert,wolf,johnston,mathewprr}.\\ 
From the crystal structure, it is apparent that our system has chains in the $a$-direction that are isolated in both the $b$ as well as the $c$-directions (refer to Figs. \ref{fig:alonga} and \ref{fig:alongbc} and the discussions therein). Since the low temperature peak in the dc magnetic susceptibility is indicative of low dimensional magnetism and the crystal structure indicates a one dimensional spin structure, we attempted to fit the magnetic data to the Bonner-Fisher model, the celebtrated model that describes the magnetism of a one-dimensional spin 1/2 Heisenberg antiferromagnet \cite{bf}. The Hamiltonian for a Heisenberg spin system governed by nearest neighbour interaction of localised quantum spins is given by: 
\begin{equation}
H = J \sum_{i}^{N-1}\overrightarrow{S_{i}} \cdot \overrightarrow{S}_{i+1}
\label{eqn:Hamiltonian}
\end{equation} 
where  $\overrightarrow{S_{i}}$ represents the spin operator on the $i$th magnetic site (Cu$^{2+}$ sites) and $J$ is the exchange coupling constant.\\ 
So, red solid curve in Figs. \ref{fig:M-T} (a) and (b) is a fit to the expression:
\begin{equation}
\chi(T) = \chi_0 + \chi_1 + \chi_{BF}(T)
\label{eqn:susceptibility}
\end{equation} 
where $\chi_0$ is a small positive constant to account for uncoupled spins, $\chi_1$ corresponds to the diamagnetic contributions from closed atomic shells of CuD estimated from Pascal's table\cite{pascal}, and $\chi_{BF}$(T) is the magnetic susceptibility estimated from the Bonner-Fisher model \cite{bf}:
\begin{equation}
\chi_{BF} = \frac{Ng^2\mu^2_B}{k_{B}T}\frac{(0.25+0.074795x+0.075235x^2)}{(1.0+0.9931x+0.172135x^2+0.757825x^3)}   
\label{eqn:Bonner-Fischer}   
\end{equation}
where symbols have their usual meaning and $x$ = $J/$k$_B$T.
From the Pascals table,$\chi_1$ was found to be $-214.46\times10^{-6}$ cm$^3$/mol. From the fit to equation \ref{eqn:Bonner-Fischer}, $\chi_0$ was obtained as $456.7\times10^{-6}$ cm$^3$/mol and the exchange coupling constant $J/$k$_B$ as 1.23 K.
\begin{figure}
	\begin{center}
		\includegraphics[width=1\textwidth]{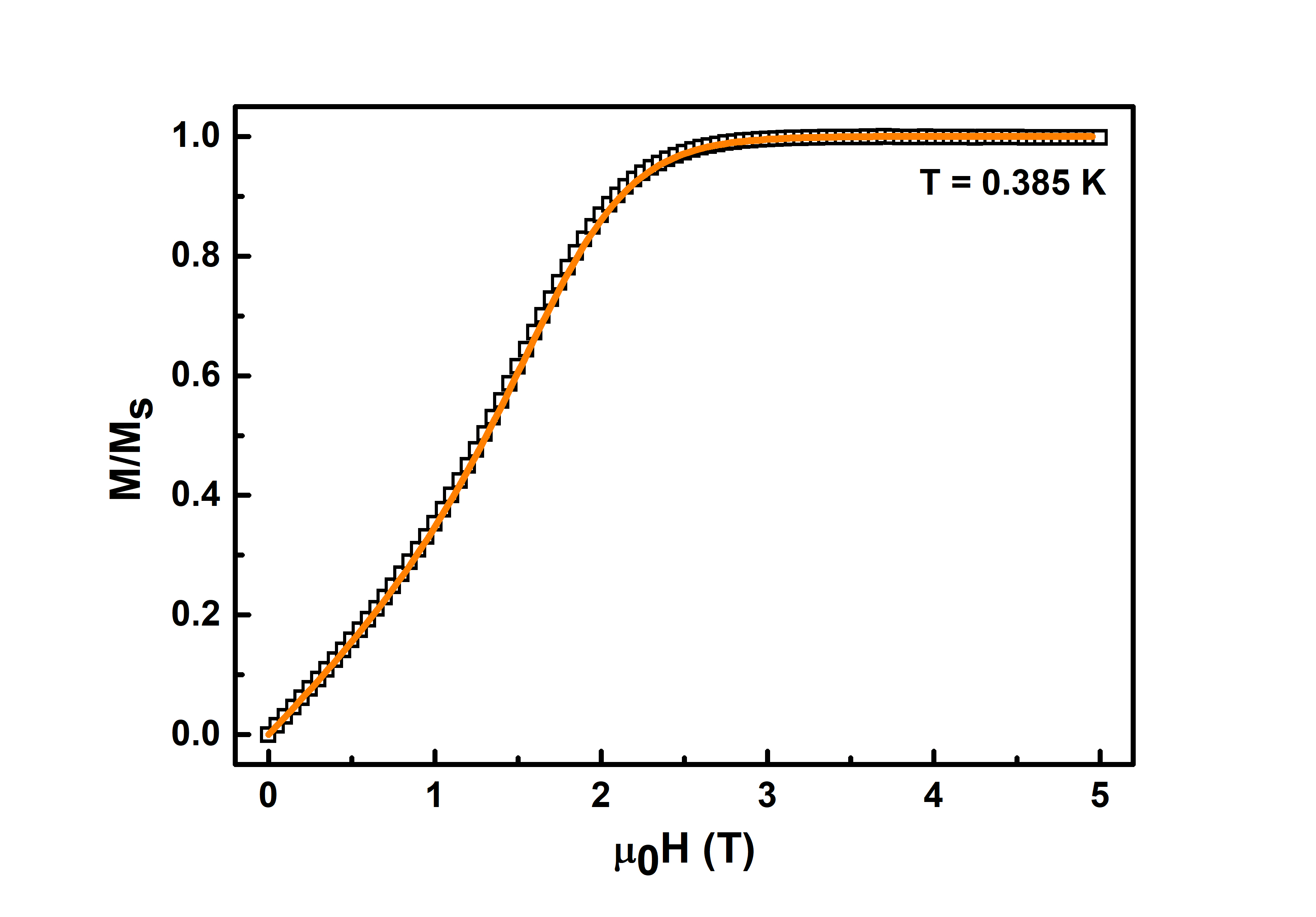}
		\caption{(colour online) Open black squares represent the magnetic field variation of dc magnetisation normalised with respect to the saturation magnetisation M$_s$ measured at a temperature of 0.385 K in C$_5$H$_7$N$_2$)$_2$[Cu(C$_2$O$_4$)$_2$].2H$_2$O. Orange solid curve is the fit to quantum Monte Carlo simuation.}
		\label{fig:M-H}
	\end{center}
\end{figure}
The maximum in the magnetic susceptibility of a spin 1/2 antiferromagnetic Heisenberg chain is expected to occur at a temperature T$_p$ = 0.640851.$|J_{1D}|/k_B$T \cite{johnston,bf}. From the obtained value of $J/k_B$ as 1.23 K, the maximum temperature T$_p$ was obtained as 0.788 K, which is exactly the same as that obtained experimentally, thus proving that our system is an excellent realisation of a spin 1$/$2 antiferromagnetic Heisenberg chain. From the fit, the axial-$g$ tensor's value was obtained as $2.029$ and the goodness of fit as $1.4\times10^{-8}$. The low value of the goodness of the fit indicates that our system is a good realisation of a spin 1$/$2 Heisenberg antiferromagnetic chain.\\
In order to understand the temperatures where antiferromagnetic correlations set-in in our system, we plotted the temperature variation of $\chi$(T)T as shown in Fig. \ref{fig:M-T} (c). It can be seen that  $\chi$(T)T is temperature independent till $\sim$ 28.5 K below which it starts to decrease sharply. This temperature is marked as T$_a$ and is the indication of the start of significant antiferromagnetic correlations between the magnetic Cu$^{2+}$ centres in the system.\\
From the above discussions, it is apparent that CuD is an excellent realisation of a spin 1$/$2 antiferromagnetic Heisenberg chain. To further confirm this nature, we measured the magnetic field dependence of magnetisation, $M$, of CuD at a temperature of 0.385 K as shown in Fig. \ref{fig:M-H}. From the figure, one can observe that the magnetisation steadily increases with an increase in the value of applied magnetic field. This happens till the saturation field H$_s$, at which the magnetisation saturates. In Fig. \ref{fig:M-H}, $M$ is plotted normalised to the value of the magnetisation at the saturation field, namely, M$_s$.\\
For a spin 1$/$2 antiferromagnetic Heisenberg chain, the relation between the exchange coupling constant J$/$k$_B$ and the saturation field H$_s$ is given as:
\begin{equation}
g \mu_B H_s = 2J/k_B
\label{eqn:Hs}
\end{equation}
where $\mu_B$ is the Bohr magneton. From equation \ref{eqn:Hs}, a H$_s$ value of 1.73 T is obtained. In order to confirm the obtained value of H$_s$, we performed a quantum Monte Carlo simulation on the magnetisation data by inputting J$/$k$_B = 1.23$ K and g = $2.029$, obtained from the Bonner-Fisher fit above. The M-H curve obtained from the simulation is plotted as solid orange curve in Fig. \ref{fig:M-H}. As can be seen from the figure, the matching between the simulated curve and the experimentally obtained data is excellent, thus proving that our facile system CuD is an excellent representation of a spin 1$/$2 antiferromagnetic Heisenberg chain. The saturation field, H$_s$, obtained from the simulation is 1.75 T, in close conformity to the 1.73 T value obtained from equation \ref{eqn:Hs}.     
\section{Conclusion}
To conclude, we have synthesized the single crystals of a facile spin 1$/$2 Heisenberg antiferromagnet that is a metal organic coordination compound, Bis(4-aminopyridinium) bis(oxalato)cuprate(II) dihydrate. The crystals were grown by the technique of liquid$/$liquid diffusion that generated large sized crystals of the order of few mm with 100 $\%$ yield. Single crystal XRD measurements revealed that the crystals stabilised in the monoclinic crystal structure with P 21$/$c space group. Density functional theory was employed to generate the optimsed geometry that matched very well with that obtained from SCXRD. The structure was found to comprise corner sharing oxygen atoms of the CuO$_6$ octahedra resulting in Cu-Cu chains in the $a$-direction that were very well isolated in the $b$ and $c$ directions. Presence of CuO$_6$ octahedra was found in the experimentally and theoretically obtained FTIR spectra. dc magnetic susceptibility and magnetisation measurements confirmed the one-dimensional character of the Cu-Cu chains where a low temperature peak at 0.78 K was found. The experimentally obtained $\chi$-T curve was found to be fitted to the Bonner-Fisher model very well that revealed the values of exchange coupling constant J$/$k$_B$ as 1.23 K. The corresponding saturation field was predicted to be 1.5 T that was confirmed from field dependent magnetisation measurements and Quantum Monte Carlo simulations, thus, affirming that our facile system is a very good representation of a spin-1$/$2 antiferromagnetic Heisenberg chain.

\end{document}